\documentclass{jfm}
\usepackage{graphicx}
\usepackage{graphics}
\usepackage{mathrsfs}
\usepackage{amsmath}
\usepackage{subfigure}

\shorttitle{Enhanced heat transport in RBC with rod-like expandable particles}
\shortauthor{S.-Y. Hu, K.-Z. Wang, L.-B. Jia, J.-Q. Zhong and J. Zhang}

\title{Enhanced heat transport in thermal convection with suspensions of rod-like expandable particles}

\author{Shi-Yuan Hu\aff{1,2,3},
  Kai-Zhe Wang\aff{1,2,3},
  Lai-Bing Jia\aff{1,4},
  Jin-Qiang Zhong\aff{5},
 \and Jun Zhang\aff{1,2,3}
 \corresp{\email{jun@cims.nyu.edu}}}

\affiliation{\aff{1}NYU-ECNU Joint Research Institute of Physics at NYU Shanghai, Shanghai 200062, China
\aff{2}Applied Math Lab, Courant Institute, New York University, New York, NY 10012, USA
\aff{3}Department of Physics, New York University, New York, NY 10003, USA
\aff{4}Department of Naval Architecture, Ocean and Marine Engineering, University of Strathclyde, G4 0LZ Glasgow, UK
\aff{5}School of Physics Science and Engineering, Tongji University, Shanghai 200092, China
}

\begin{document}

\maketitle

\begin{abstract}
Thermal convection of fluid is a more efficient way than diffusion to carry heat from hot sources to cold places. Here, we experimentally study the Rayleigh-B\'enard convection of aqueous glycerol solution in a cubic cell with suspensions of rod-like particles made of polydimethylsiloxane (PDMS). The particles are inertial due to their large thermal expansion coefficient and finite sizes. The thermal expansion coefficient of the particles is three times larger than that of the background fluid. This contrast makes the suspended particles lighter than the local fluid in hot regions and heavier in cold regions. The heat transport is enhanced at relatively large Rayleigh number ($\Ray$) but reduced at small $\Ray$. We demonstrate that the increase of Nusselt number arises from the particle-boundary layer interactions: the particles act as ``active'' mixers of the flow and temperature fields across the boundary layers. 
\end{abstract}


\section{Introduction}\label{sec1}
Convection driven by a temperature difference is omnipresent in nature, such as convection in oceans~\citep{Marshall99}, in the atmosphere~\citep{Hartmann01}, inside the Earth~\citep{Jones11,Cardin94}, and in the outer layer of the Sun~\citep{Cattaneo03}. It is also relevant to numerous technological applications ranging from the cooling of electronic devices to the ventilation in buildings~\citep{Incropera99,Linden99}. In many cases, thermal convection is coupled to other physical processes, such as phase changes~\citep{Kim09}, transport and dispersion of inertial particles~\citep{Shaw03,Ackerman04}, leading to complex and interesting dynamics.

Rayleigh-B\'enard convection (RBC) has been extensively studied as a simplified paradigm for thermal convection~\citep{Castaing89,Ahlers09,Chilla12,Xia13}. Buoyancy driven flow in RBC is induced by heating a fluid layer from below and cooling it from above. The system is characterized by two control parameters: the Rayleigh number $\Ray$ and Prandtl number $\Pran$, defined as 
\begin{equation}\label{eq2}
\Ray = \alpha gH^3 \Delta T/\nu \kappa \quad \text{and} \quad \Pran = \nu/\kappa,
\end{equation}
respectively. Here, $\Delta T$ is the temperature difference applied across the fluid layer, $H$ is the height of the fluid layer, and $g$ is the gravitational acceleration. The parameters of the fluid, $\alpha$, $\nu$, and $\kappa$, are the thermal expansion coefficient, kinematic viscosity, and thermal diffusivity, respectively. The heat transport by the convective flow is characterized by the dimensionless Nusselt number, 
\begin{equation}\label{eq1}
\Nu = QH/\chi \Delta T,
\end{equation}
where $Q$ is the heat flux (power per unit area) through the convection cell and $\chi$ is the thermal conductivity of the fluid. The dependence of $\Nu$ on $\Ray$ and $\Pran$, $\Nu = f(\Ray, \Pran)$, has been the focus of numerous experimental and theoretical studies~\citep[for review see e.g.,][]{Ahlers09}.

An important question in thermal convection is how to modify the fluid structures, change the heating-passing mechanisms, and perhaps enhance the heat transport efficiency. Various approaches have been proposed and tested, such as creating roughness on the top and bottom plates~\citep{Shen96,Du98}, exploiting multiphase working fluid~\citep{Zhong09,Lakkaraju13,Dabiri15,Guzman16,Wang19,Liu21}, imposing spatial confinement~\citep{Huang13}, partitioning the convection cell~\citep{Bao15}, and manipulating the coherent structure~\citep{Chong17}. 

Another potential approach to tune heat transport in RBC is to couple thermal convection with inertial particles. In particle-laden flows, the particle inertia, i.e., lagging in responses to the changes in the fluid flows, can arise from the density mismatch between particle and the local fluid or particle finite size~\citep{Ouellette08,Cartwright10}. Previous studies have focused on micron- or millimeter-scale particles. Particles may settle out of the fluid when there is a mismatch between the densities of the particles and the background fluid. A turnover of the system dynamics and resuspension of settling particles have been identified beyond a critical particle concentration~\citep{Koyaguchi90}. 
The settling particles may form porous layers that reduce the heat transport~\citep{Joshi16}. 
The distribution of thermally expandable point-like particles in RBC has also been studied numerically in the ``soft turbulent'' regime, with the assumption that the presence of the particles does not modify the flow~\citep{Alards19}. However, to the best of our knowledge, no enhancement of $\Nu$ has been reported in turbulent RBC with solid inertia particles. When the particle size is much larger than the Kolmogorov length scale~\citep{Kolmogorov41}, the effect of the motion and geometry of particles on fluid flows must be considered, and whether $\Nu$ can be enhanced is still an open question.

The study of thermal convection with large inertial particles has been challenging both numerically and experimentally. In numerical simulations, one has to properly resolve the fluid-particle coupling, particle-particle and particle-wall collisions~\citep{Balachandar10,Mathai20}. In experiments, as particle size increases, gravitational settling due to density mismatch becomes more pronounced. Therefore, the uniformity of the particle density and the close density match to the background fluid are essential to ensure and maintain stable suspensions. In contrast to infinitesimal passive tracer particles commonly used to map the velocity field, inertial particles may depart from the local flow and show various behaviors, such as mixing, separation, and aggregation~\citep{Fung03,Saw08,Sudharsan16}. 

Here, we experimentally study the effect of inertial rod-like particles suspended in convecting flows on the overall heat transport. The particle lengths are centimeter-scale, much larger than the Kolmogorov length scale and the thickness of the thermal boundary layers (BLs). They are made of polydimethylsiloxane (PDMS). Besides the large size, the particle inertia is further enhanced by its large thermal expansion coefficient (see table~\ref{table1}): the particles are lighter than the fluid in hot regions and heavier than the fluid in cold regions. For the first time, we observe an increase in $\Nu$ in turbulent RBC with suspensions of solid inertial particles due to the interactions between particles and the background flows, especially within the BLs.

\section{Experimental procedures}\label{sec2}
\begin{figure}
\centering
\includegraphics[bb= 0 3 353 145, scale=1,draft=false]{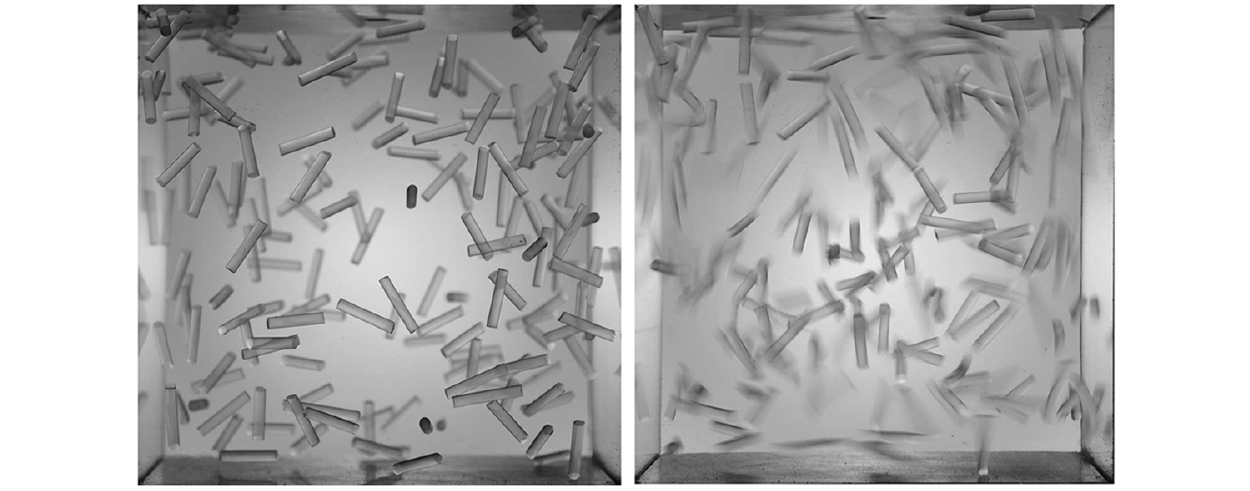}
\caption{The suspended PDMS particles of length 2.5 cm and diameter 0.45 cm with volume fraction $\phi = 1\%$, interact with the background turbulent flows in Rayleigh-B\'enard convection at $\Ray \approx 2.1\times 10^{9}$. Left: 1/100 sec exposure photo shows how particles are randomly distributed in the convection cell, and reach the top and bottom plates occasionally. Right: 1 sec long exposure photo shows different degrees of motion blurs that reflect the turbulent background flows.}
\label{fig1}
\end{figure}
\begin{table}
  \begin{center}
  \begin{tabular}{p{4.cm}p{3.7cm}p{3.5cm}}
        & Aqueous glycerol solution & PDMS particles\\
       \hline
       Density $\rho$ (at 27$^{\circ}$C) & $1.03\times 10^{3}$ kg/m$^{3}$ & $1.03\times 10^{3}$ kg/m$^{3}$\\
       Thermal expansion coef. $\alpha$ & 3.03$\times10^{-4}$ K$^{-1}$ & 9.3$\times10^{-4}$ K$^{-1}$\\
       Thermal conductivity $\chi$ & 0.54 W/(m$\cdot$K) & 0.2 W/(m$\cdot$K)\\
       Specific heat capacity $c_p$ & 3.8$\times10^3$ J/(kg$\cdot$K) & 1.4$\times10^3$ J/(kg$\cdot$K) \\
       Prandtl number $\Pran$ & 8.7 & --\\
       Particle length $L$ & -- & 1.25, 2.50 cm\\
       Particle diameter $d$ & -- & 0.30, 0.45 cm\\ 
       \hline
  \end{tabular}
  \caption{Parameters of the working fluid and PDMS particles. Densities and the particle length are measured in the experiments. Other parameters are from empirical equations or interpolated using known experimental values~\citep{Cheng08,Volk18,Mark99}.}
  \label{table1}
  \end{center}
\end{table}

Our experiments have been conducted in a cubic cell of size $20\times20\times19.6$ cm$^{3}$ with the cell height $H=19.6$ cm. The sidewalls consist of four glass plates of 0.5 cm in thickness. The top cooling plate and the bottom heating plate are made of surface-anodized aluminum. The top plate is cooled by passing temperature-regulated water through its internal grooves. The bottom plate is heated by a film heater at constant power. The temperatures of the top plate and bottom plate are monitored by thermistors embedded within them at roughly 1 mm away from the fluid contact surfaces, which leads to short thermal response time and allows the thermistors to perceive the temperature fluctuations in the boundary layers~\citep{Verzicco04}. The bulk temperature $T_c$ is measured by two thermistors extended into the center of convection cell.

The suspended particles are made of PDMS, which has been widely used in microfluidics~\cite[for review, see][]{Sackmann14} due to its advantageous properties, such as chemical stability and mechanical flexibility after curing. However, the properties relevant to our purposes are different. First, its density after curing is about 3\% within that of water. A stable suspension of particles, which is crucial to our experiments, is achieved by tuning the background fluid: a dilution of glycerol in water. The density of the working fluid is carefully adjusted such that the particles are neutrally buoyant at 27$^\circ$C: particles can suspend for hours even in quiescent fluid. The working fluid is then used for all the subsequent experiments, which are conducted in the range $2^{\circ}\mathrm{C}\lesssim \Delta T \lesssim 32^{\circ}\mathrm{C}$ and $3\times 10^{8}\lesssim \Ray \lesssim 4\times 10^{9}$. The $\Ray$ range is about 2--3 orders of magnitude larger than the $\Ray$ values in~\cite{Alards19}, and about one order of magnitude smaller than that in~\cite{Wang19} and~\cite{Joshi16}. More importantly, PDMS has a large thermal expansion coefficient that is more than three times that of the working fluid (table~\ref{table1}), making the particle thermally responsive and more active than the background fluid. 

All PDMS particles are handmade. After many trials, we found that adopting rod-like geometry yields high precision and high quality. To make these particles, the PDMS fluid and the curing agent (SYLGARD 184 Silicone Elastomer Kit, Dow Corning) are first mixed at a 10:1 volumetric ratio and degassed for 30 min by a vacuum pump. The mixture is then poured into molding tubes of desired diameters, and cured at room temperature for 48 hours. The cured PDMS rods are then cut into desired lengths. Making spherical particles (a simpler geometry than rod) using a similar procedure turns out to be impractical, especially when making thousands of identical particles. 

Figure~\ref{fig1} presents the photos of suspended particles in turbulent convection with particle volume fraction $\phi = 1\%$, where $\phi$ is defined as the ratio of the total volume of particles to the volume of the convection cell. The heating power and cooling temperature are pre-adjusted such that the bulk temperature $T_{\mathrm{c}} = 27 \pm 0.10^\circ$C for no-particle convection experiments. For experiments with particles, two control modes are used. The first one is constant-$T_\mathrm{c}$ mode: for different particle volume fractions $\phi$, the heat flux $Q$ is regulated through a feedback control, and once the system reaches dynamic equilibrium with $T_{\mathrm{c}} = 27\pm 0.15^{\circ}$C, $Q$ is fixed and no longer changes over time. The second mode is constant-$Q$, in which the same $Q$ is used for different values of $\phi$. When measuring $\Nu$ in response to the effects of added particles, constant-$T_\mathrm{c}$ mode is used to keep rigorously the bottom-top symmetry of the particle distributions in the convection cell. Constant-$Q$ mode is mainly used to measure temperature fluctuations. 
%
\section{Results and Discussions}\label{sec3}

\begin{figure}
\centering
\includegraphics[bb= 0 5 365 105, scale=1,draft=false]{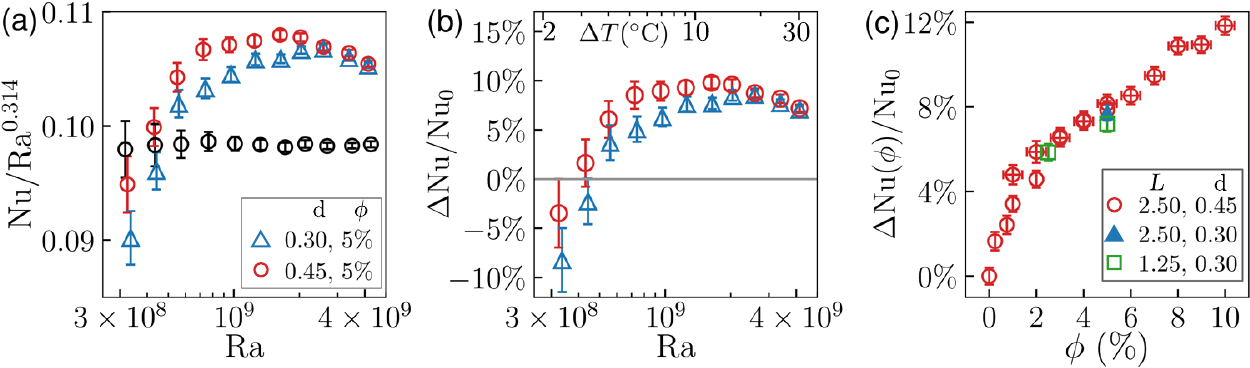}
\caption{With constant-$T_{\mathrm{c}}$ mode, enhancement of heat transport with PDMS particles. (a) $\Nu/\Ray^{0.314}$ as a function of $\Ray$ in semi-log scale. Gray filled circles show data measured without particles; other symbols represent measurements with particles of different diameter $d$. Error bars represent uncertainties due to temperature fluctuations and heat loss to the ambient air. (b) Data from (a), plotted as relative increase of $\Nu$, $\Delta \Nu/\Nu_0$ (left axis), as a function of $\Ray$. Upper $x$ axis show $\Delta T$ without particles. (c) $\Delta \Nu/\Nu_0$ as a function of $\phi$ at $\Ray \approx 3.4\times 10^{9}$ for $d=0.45$ cm and $L=2.50$ cm (red circles). Blue triangle shows data for $d=0.30$ cm and $L=2.50$ cm at $\phi=5.0\%$; green squares show data for $d=0.30$ cm and $L=1.25$ cm at $\phi=2.5\%$ and $5.0\%$. Particle aspect ratio plays a less important role on $Nu$ than $\phi$, as can be seen by comparing rods of different geometries.}
\label{fig2}
\end{figure}

In figure~\ref{fig2}(a), we show the compensated Nusselt number, $\Nu/\Ray^{0.314}$, as a function of $\Ray$ in a semi-log scale. The measured values of Nusselt number $\Nu_0$ from no-particle experiments follow a power law with $\Ray$, and the scaling exponent agrees with previous experiments~\cite[see, e.g.,][]{Niemela00}. We then measure $\Nu$ with the suspensions of particles of different diameters and compute the relative changes of $\Nu$, $\Delta \Nu/\Nu_0 = (\Nu-\Nu_0)/\Nu_0$. As shown in figure~\ref{fig2}(a) and figure~\ref{fig2}(b), $\Delta \Nu/\Nu_0$ depends strongly on $\Ray$, and an optimal value of $\Ray$ exists, around which the $\Nu$ enhancement is most obvious. For particles of diameter 0.30 cm, the maximum increase of $\Nu$ is achieved around $\Ray\approx 2\times 10^{9}$; for particles of a larger diameter 0.45 cm, the maximum is located at a slightly lower $\Ray$. As $\Ray$ becomes larger than the optimal values, $\Delta \Nu/\Nu_0$ decreases slowly. However, lowering the values of $\Ray$, $\Delta \Nu/\Nu_0$ decreases steeply and even turns negative at sufficiently small $\Ray$. Figure~\ref{fig2}(c) shows that $\Delta\Nu(\phi)$ increases faster as $\phi$ is increased from 0 but slows down for larger values of $\phi$. It is evident that the addition of the passive inertia particles increases the heat-transport efficiency in RBC. For higher values of $\phi$, the particles are likely to get trapped at the corners of the convection cell and form porous layers covering the top or bottom plates, making the mixture no longer a mobile suspension.
%
\begin{figure}
\centering
\includegraphics[bb= 0 5 365 92, scale=1,draft=false]{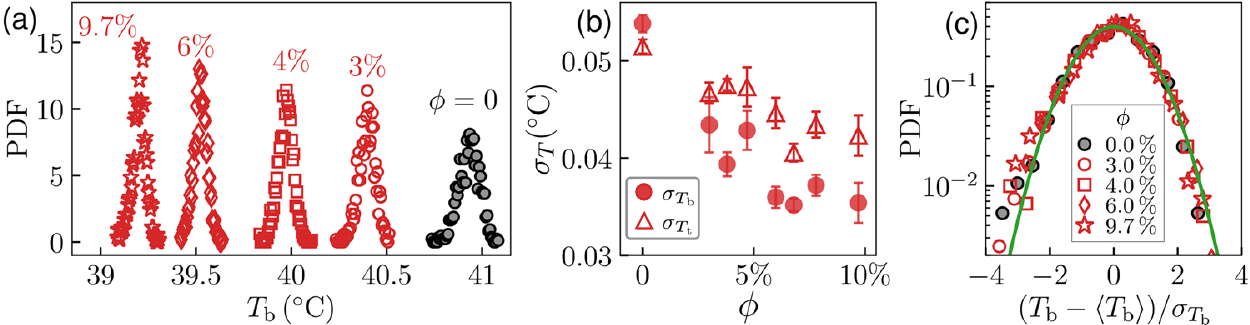}
\caption{With constant-$Q$ mode, $T_{\mathrm{t}}$ remains within $10.5 \pm 0.1^{\circ}$C, (a) probability distribution function (PDF) of $T_\mathrm{b}$ for $\Ray\approx 3.9\text{--}4.2\times 10^9$ for no-particle experiments ($\phi=0$, gray solid circles) and experiments with particles of diameter 0.30 cm and length 2.50 cm (red symbols) at different values of $\phi$; blue symbols show particles of diameter 0.30 cm and length 1.25 cm. (b) the standard deviation $\sigma_T$ of the bottom temperature $T_\mathrm{b}$ (red circles) and top temperature $T_\mathrm{t}$ (blue triangles) as a function of $\phi$; (c) PDF of normalized $T_\mathrm{b}$ for different values of $\phi$ in semi-log scale. The green curve show a standard Gaussian distribution. Statistics are obtained from time series of 2400 sec at a sampling rate of 0.5 Hz.}
\label{fig3}
\end{figure}

The Stokes number $St$ of the rod-like particles can be estimated as $St \sim \bar{\rho}d^2 H^{-1} h(\lambda) v_f/\nu$, where $\bar{\rho}$ is the density ratio of particle to fluid, $\bar{\rho} = \rho_p/\rho_f$, $v_f$ is the characteristic velocity of the large scale circulation, and $h(\lambda)$ is a function of the particle aspect ratio $\lambda$ with $\lambda=L/d$~\citep{Voth17}. We estimate $v_f$ as $v_f\approx (0.2\nu/H) \sqrt{\Ray/\Pran}$, where the prefactor 0.2 accounts for the correction at large \Pran~\citep{Silano10}. For the $Ra$ range in our experiments, $v_f \approx 0.58-2.6$ cm/s, and $St \approx 0.04-0.16$ for particles of $L = 2.5$ cm and $d=0.3$ cm, which is orders of magnitude larger than those reported in previous experiments and simulations~\citep{Ouellette08,Lopez17}, indicating finite inertia effect afforded by the large particle size and deviations of particle trajectories from fluid flows. To see the effect of particle aspect ratio, we manually cut each particle of $d=0.30$ cm in half. As shown in figure~\ref{fig2}(c), for the same volume fraction at $\phi=5\%$, $\Delta Nu$ of particles of $L=1.25$ cm (right green square) is nearly the same as that of particles of double length (blue triangle); for the same number of particles, the $Nu$ of particles of $L=1.25$ cm at $\phi=2.5\%$ (left green square) is slightly smaller than that of particles of double length at $\phi=5.0\%$ (blue triangle). Given the current large particle sizes at centimeter scale, our data (within experimental accuracy) show that the particle aspect ratio has minor effect on $Nu$. We speculate that the aspect ratio are more important for thinner particles of $St \ll 1$, and more experiments are needed in the future to further explore the effect of elongated shape of particles. 

Figure~\ref{fig3}(a) shows the probability distribution function of the bottom plate temperature $T_{\mathrm{b}}$ around $\Ray$ values where particles enhance $\Nu$. With constant-$Q$ mode, i.e., the heating power of the bottom plate is fixed to be the same as that used in the no-particle experiment, $T_{\mathrm{b}}$ decreases as $\phi$ is increased, indicating a decrease of $\Delta T$ and thus an increase of $\Nu$. This is in line with the increase of $\Nu$ shown in figure~\ref{fig2}(c) under constant-$T_{\mathrm{c}}$ mode, in which $\Delta T$ remains relatively fixed within $\pm 0.4^{\circ}$C and the heating power of the bottom plate increases as $\phi$ is increased. Meanwhile, as shown in figure~\ref{fig3}(b), the standard deviation $\sigma_T$, which characterizes the magnitude of the temperature fluctuations, decreases as $\phi$ is increased for both top and bottom plates. For different values of $\phi$, the normalized bottom temperatures, $(T_{\mathrm{b}}-\langle T_{\mathrm{b}} \rangle)/\sigma_{T_{\mathrm{b}}}$, collapse onto each other and closely resemble a Gaussian distribution (figure~\ref{fig3}(c)). It is well accepted that thermal BLs impose the most resistance on heat transport and dominantly determines $\Nu$~\cite[see, e.g.,][]{Castaing89}. Much of the heat flux is carried by the thermal plumes, which are emitted from the thermal BLs and contrast the background flow in temperature and momentum~\citep{Zocchi90,Kadanoff01}. The decrease in $\sigma_T$ observed in figure~\ref{fig3}(b) suggests that particles act as mixers of the flow and temperature fields near the BLs. The large temperature fluctuations caused by the emission and arrival of thermal plumes are partially smoothed out by the motion of the particles.

\begin{figure}
\centering
\includegraphics[bb= 0 2 360 70, scale=1,draft=false]{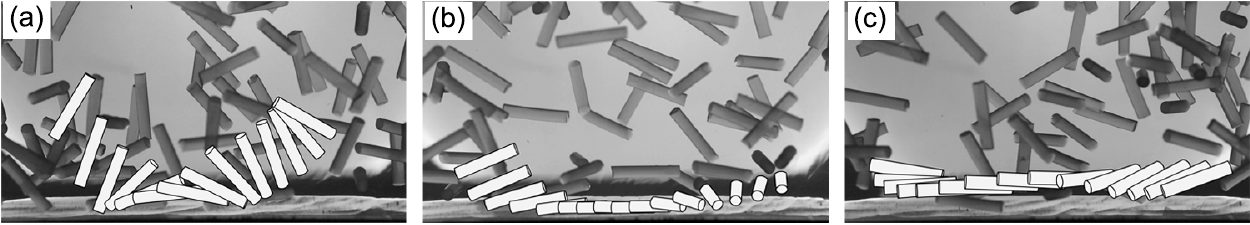}
\caption{Three typical examples of how particles of length 2.5 cm and diameter 0.45 cm interact with bottom BL at $\Ray \approx 2.1\times 10^{9}$. Successive snapshots (white cylinders) showing particles moving from left to right as extracted from the video with a time interval of 1.0 sec for (a) and (c) and 0.7 sec for (b). The background pictures are taken from the last frame. (a) A particle briefly collides twice with the BL and returns to the bulk. (b) Particle slides along the BL. (c) Particle hovers above the BL.}
\label{fig4}
\end{figure}
The path and orientation of each particle seem to be complex, as depicted in figure~\ref{fig1}, more so are the interactions between particles and BLs. Several forms of interactions between particles and BLs are observed (see figure~\ref{fig4} and the supplementary movie). The particles can collide with the top and bottom plates and then return to the bulk (figure~\ref{fig4}(a)); they may also slide along the top and bottom plates, within or close to the BLs (figure~\ref{fig4}(b) and \ref{fig4}(c)). 

\begin{figure}
\centering
\includegraphics[bb= 0 2 370 110, scale=1,draft=false]{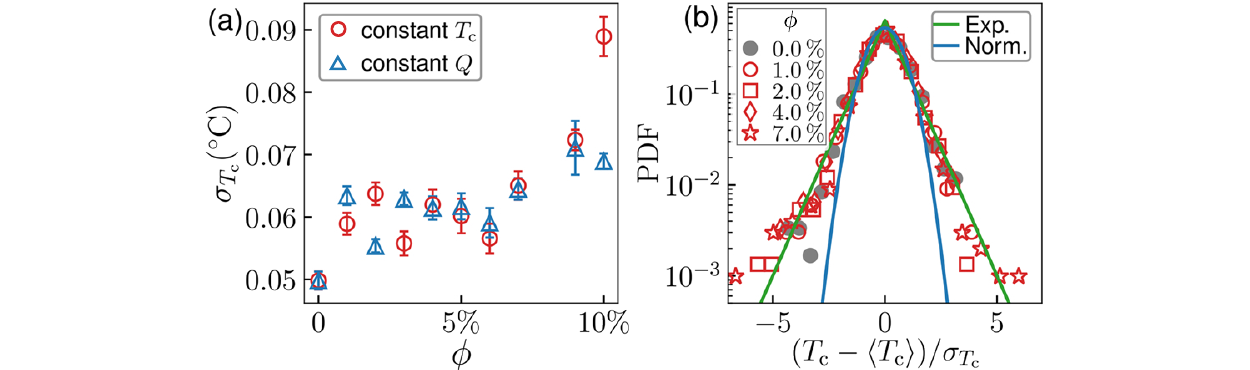}
\caption{Statistics of the bulk temperature $T_{\mathrm{c}}$ obtained from time series of 2400 sec at a sampling rate of 0.5 Hz. (a) $\sigma_{T_\mathrm{c}}$ as a function of $\phi$ with constant-$T_{\mathrm{c}}$ (red circles) and constant-$Q$ modes (blue triangles). (b) Normalized distributions of $T_\mathrm{c}$ for different values of $\phi$ in semi-log scale with the best-fit exponential distribution (green curve), which has a better agreement with the experimental data than the best-fit normal distribution (blue curve).}
\label{fig5}
\end{figure}

The estimated thickness of thermal BL, $\delta = H/2 \Nu \approx 0.1 \text{--} 0.2\ \text{cm}$, is close to the particle radius, and the particles are exposed to large temperature variations when getting close to the thermal BLs. Due to the large thermal expansion coefficient of PDMS, the particles become lighter than local fluid when in contact with the hot BL, and heavier than local fluid when in contact with the cold BL. At the largest value of Ra, the maximum and minimum density ratio $\bar{\rho}$ is about $1.01$ and $0.99$, respectively, taking into account both the thermal expansion coefficients of the particles and fluid. Despite only one percent density variation, the particles may significantly deviate from fluid flows due to their large sizes. We estimate the buoyancy based settling/rising velocity in quiescent fluid as $v_g \approx \left[|\bar{\rho}-1|\times g L\right]^{1/2}$~\citep{Auguste18}. When particle length $L = 2.50$ cm, $v_g \approx 1.5-5.0$ cm/s for the $Ra$ range in our experiments. If the particle diameter is taken as the characteristic scale, $v_g \approx 0.55-1.7$ cm/s. Both estimates are comparable to $v_f$, and therefore, the particles become more ``active'' than the local fluid within BL, making them more efficient in carrying fluid and heat into the bulk. As a result, the standard deviation of the bulk temperature $\sigma_{T_{\mathrm{c}}}$ increases with $\phi$ under both constant-$T_{\mathrm{c}}$ and constant-$Q$ modes (figure~\ref{fig5}(a)). As shown in figure~\ref{fig5}b, the normalized $T_{\mathrm{c}}$ displays exponential distributions with relatively large deviations from the means~\cite[see, e.g.,][]{Castaing89}. The enhancement of fluctuations in incident turbulence has been observed due to the agitation of bubbles~\citep{Almeras17}. In RBC with vapour-bubble nucleation, the temperature fluctuations are decreased probably due to vapour bubbles absorbing and releasing heat~\citep{Lakkaraju14,Guzman16}. This effect is negligible in our experiments since no latent heat (no phase transition) is involved and the specific heat capacity of PDMS is much smaller than that of the background fluid. It has also been observed that the bulk temperature fluctuations are increased in convection cells with rough surfaces compared with smooth surfaces, along with an increase in the heat transport efficiency~\citep{Du01}.

To make particles more ``active'' than the background fluid, experiencing large $\Delta T$ and having sufficient time for particles to thermally adapt to $\Delta T$ are needed. The latter is determined by two time scales. The first one is the thermal response time of the particles, $\tau_p \sim d^{2}/4\kappa_p$, where $\kappa_p$ is the thermal diffusivity of PDMS. The other one is the characteristic time during which the particles remain close to the BLs, $\tau_f$, which decreases as $\Ray$ is increased. We estimate $\tau_f$ as $\tau_f \approx H/v_f$. For the $\Ray$ range in our experiments, the ratio of the two characteristic times $\bar{\tau}=\tau_p/\tau_f\approx$ 0.5--2.4. At large $\Ray$, $\bar{\tau}>1$, although $\Delta T$ is large, $\tau_f<\tau_p$ and the particles have insufficient time to warm up or cool down in response to the high/low temperatures of the bottom/top BLs before returning to the bulk. At small $\Ray$, $\bar{\tau}<1$ and $\tau_f>\tau_p$, but $\Delta T$ is small, leading to relatively small variations in the densities of the particles, i.e., the particles are not ``active'' enough even after a full thermal relaxation to the temperatures of the BLs. Indeed, we observe at small $\Ray$, that the particles occasionally accumulate and stay near the bottom and top plates for relatively long times, blocking the thermal plumes and thus decreasing $\Nu$. At intermediate $\Ray$, where $\Delta T$ is relatively large and $\tau_f \approx \tau_p$, $\Delta \Nu/\Nu_0$ reaches the largest value. In this regime, a particle gains enough density difference and returns to the bulk with some fluid dragged by it. It might be able to intrude the BL of the other side, switch to another temperature and density extreme, and repeat this process. The intrusion of the particles to the thermal BLs is probably the most responsible mechanism at play in enhancing the overall heat-transport efficiency, reflected in the increase of $\Nu$. 

\begin{figure}
\centering
\includegraphics[bb= 0 2 352 110, scale=1,draft=false]{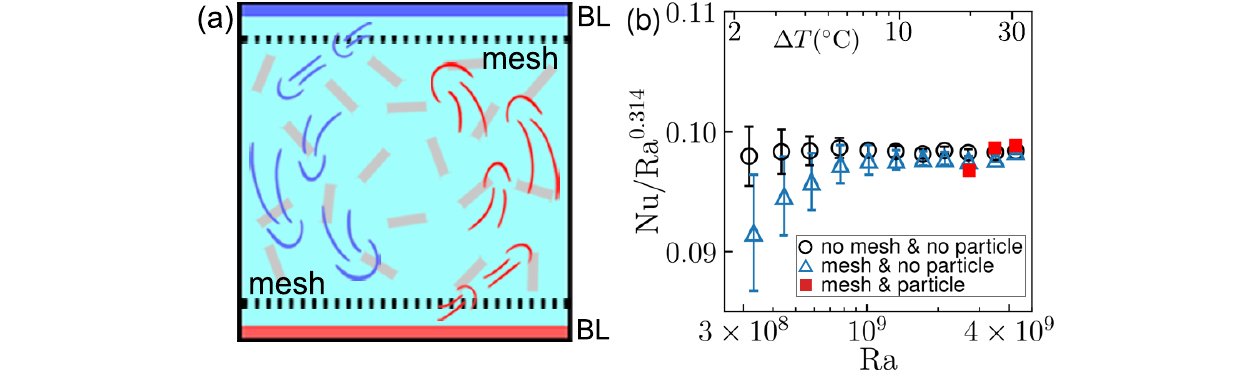}
\caption{(a) Schematic of experiment with particles and meshes, which prevent particles from getting close to the BLs. (b) $\Nu/\Ray^{0.314}$ as a function of $\Ray$ without mesh and particle (gray filled circles), with mesh and no particles (blue triangles), and with mesh and particles (red solid squares) of diameter 0.45 cm and length 2.50 cm with $\phi = 6\%$. Enhancement of $\Nu$ is no longer observed.}
\label{fig6}
\end{figure}
To directly verify that it is the particle-BL interaction that leads to the enhancement of heat transport efficiency, as illustrated in figure~\ref{fig6}(a), we mount two polythene mesh sheets outside the thermal BLs at 1 cm away from the top and bottom plates, respectively. Each mesh has about $56\%$ opening area and its thickness or wire diameter is 0.1 cm. The mesh opening has a diamond shape of size $0.3 \times 0.3$ cm, smaller than the $d=0.45$ cm particles, and thus prevents them from passing through. The mesh sheets have little effect on the heat transport at large $\Ray$ (figure~\ref{fig6}(b), blue triangles). We then add particles, which are kept within the two meshes and circulating in the bulk without interacting with the BLs. As evidenced in figure~\ref{fig6}(b) (red squares), the measured $\Nu$ with particles has a negligible difference from the $\Nu$ measured without particles. Indeed, particle-BL interaction is crucial in the enhancement of heat transport.

\section{Conclusion}\label{sec4}
\bigskip
We have experimentally studied a cubic RBC system with suspensions of inertial and expansible rod-like PDMS particles. The heat transport may be enhanced or reduced depending on the Rayleigh number. Our measurements demonstrate that the particle's large thermal expansibility has strong effects on the heat transport. Our results may shed light on a new approach to control the heat transport in thermal convection with suspensions of thermally responsive particles without modification to the classical convection system. Our results may also have implications for the transport and mixing of particles in complex flows and in confined environments. 

Engineered materials with voids filled with air, such as silicon foams, may further enhance the heat transport, since the thermal expansion coefficient of air is more than ten times that of water. Particles made of these materials are expected to gain large density variations across temperature differences in thermal convection, but density match may be difficult to achieve. In this work, we have not quantified the 3D dynamics of the particles, which could be constructed using 2D images taken by multiple cameras~\citep{Parsa12}. In convection cells with extreme aspect ratios where multiple rolls exist, the dynamics is expected to be more complex since particles may move across different convection rolls~\citep{Solomon88}. We speculate that in a confined rectangular cell (with a narrower width), the interactions between particles and boundary layers may become more frequent, and $\Nu$ may be further enhanced. 

\bigskip

{\noindent \bf Acknowledgement.} We thank M. Huang for helpful discussions and C. Sun for useful references. We also thank the anomalous reviewers for their insightful criticisms and constructive suggestions. S.Y.H and K.Z.W. gratefully acknowledge support from the MacCracken Fellowship provided by New York University. J.Z. acknowledges support from NYU Shanghai and partial support by Tamkeen under the NYU Abu Dhabi Research Institute grant CG002.
\bigskip

{\noindent \bf Declaration of Interests.} The authors report no conflict of interest.

\bibliographystyle{jfm}
\bibliography{reference}

\end{document}